\documentclass[twocolumn,prb,showpacs]{revtex4}

\usepackage{epsfig}
\usepackage{graphics}
\usepackage{graphicx}
\usepackage{dcolumn}
\usepackage{amsmath,amsfonts,amssymb,amsxtra}
\usepackage{mathrsfs}
\usepackage{bm}
\usepackage{times}

\def\MgB2{MgB$_{2}$}
\def\cm-1{cm$^{-1}$\,}
\def\cmT-1{cm$^{-1}$/T\,}
\def\E2g{$E_{2g}$}
\def\A1g{$A_{1g}$}
\def\2DS{$2\Delta_{S}^{E}$}

\def\2DA{$2\Delta^{A}$}
\def\D0{$2\Delta_{0}$}
\def\Tc{$T_c$}
\def\D6h{D$_{6h}$}

\begin{document}

\title{Anharmonicity and self-energy effects of the \E2g phonon in
\MgB2}

\author{A. Mialitsin$^{1,\dag}$}
\author{B. S. Dennis$^{1}$}
\author{N.~D.~Zhigadlo$^{2}$}
\author{J.~Karpinski$^{2}$}
\author{G.~Blumberg$^{1,*}$}

\affiliation{
$^{1}$Bell Laboratories, Alcatel-Lucent, Murray Hill, NJ 07974 \\
$^{2}$Solid~State~Physics~Lab,~ETH,~CH-8093~Z\"urich,~Switzerland\\
$^{\dag}$Rutgers University, Piscataway, NJ 08854}

\date{October 4, 2006}

\begin{abstract}

We present a Raman scattering study of the \E2g phonon anharmonicity
and of superconductivity induced self-energy effects in \MgB2
single crystals. 
We show that anharmonic two phonon decay is mainly responsible for the
unusually large linewidth of the \E2g mode. 
We observe $\sim 2.5\%$
hardening of the \E2g phonon frequency upon cooling into the 
superconducting state and estimate the electron-phonon coupling
strength associated with this renormalization.

\end{abstract}

\pacs{74.70.Ad, 74.25.Ha, 74.25.Gz, 78.30.Er}

\maketitle

High-$T_c$ superconductivity in \MgB2 is known to be promoted
mainly due to the boron layers \cite{KortusPRL86Sc_of_metallicBinMgB2},
thus the high frequency
lattice vibrations of light boron atoms beneficially increase the
electron-phonon coupling.
The \E2g Raman active in-plane boron vibrational mode contributes
significantly to
superconductivity; this fact is reflected by the Eliashberg
function $\alpha^2\,F(\omega)$ peaking in the
same frequency range where a high phononic density of states is
accounted for by Van Hove singularities of the \E2g branch in
the $\Gamma$ and $A$ points of the Brillouin zone (BZ) 
\cite{DagheroPhC,Yildirim}.
The reason the \E2g mode plays a prominent role in the
superconducting (SC)
mechanism is that the mode strongly couples to the
$\sigma$-type states of the boron plane as can be seen from the
basic geometry of the electronic configuration \cite{Choi}.

Raman spectra exhibit unusually broad linewidth of the \E2g mode
\cite{RenkerJLTP,QuiltyPRL88,Goncharov} which has been the subject of
numerous speculations. 
While high impurity scattering in earlier low
quality samples has been suggested as one of the possible reasons,
this mechanism can be readily excluded with recent high quality
single crystals. The two remaining contributions to the \E2g
phonon rapid decay are 
(i) strong electron-phonon coupling and
(ii) multiphononic decay (subsequently referred to as
\emph{anharmonicity}).
The relative importance of the
electron-phonon coupling and anharmonicity in this matter is under
intense debate. 
On one hand a density functional theory 
calculation asserts that the anharmonic contribution to
the \E2g phonon linewidth is negligible ($\sim 1$\,meV) \cite{Shukla}.
On the other hand analysis of the phonon self-energy in the long
wavelength limit shows that the $\sigma$-band contribution to the
phonon decay is vanishing \cite{Calandra}.
Thus, even when
contributions of the spectral weight of
$\alpha^2\,F(\omega)|_{\omega < \omega_{E_{2g}}}$
 to the damping of the \E2g phonon are accounted for, 
\cite{Cappelluti} the experimentally observed linewidth of
25-35\,meV at low temperatures \cite{QuiltyPRL88,RenkerJLTP} cannot 
be explained with electron-phonon coupling alone whose part in
the \E2g mode linewidth at low temperatures amounts to about
6\,meV even in such an elaborate scenario as that in Ref.
\cite{Cappelluti}.
\begin{figure*}
\includegraphics[width=2.0\columnwidth]{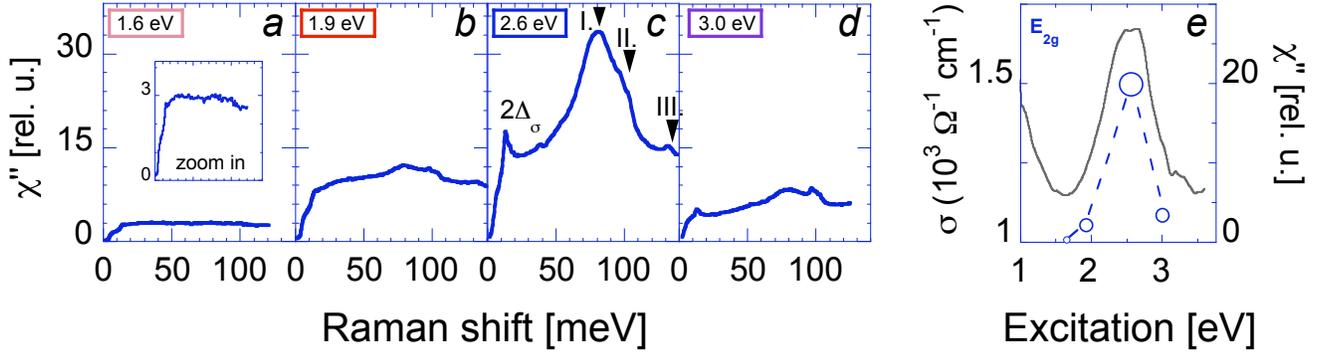}
  \caption{(Color online) 
  Raman response function from crystal ${\mathcal A}$ at
  8~K for $VH$ (\emph{a}) and $RL$ (\emph{b-d}) polarizations corresponding to
  \E2g channel for four excitation energies between 1.65 and 3.05~eV.
  The spectra exhibit a broad electronic continuum,
  a superconducting coherence peak $2\Delta_{\sigma}$,
  the \E2g boron stretching phonon mode (\emph{I}) and
  two-phonon scattering bands (\emph{II - III}).
  The comparison of $ab$-plane optical conductivity 
  (Ref.\cite{Guritanu}) and the \E2g mode intensity is
  shown in panel (\emph{e}).
  The size of the symbols is proportional to the ratios of mode to 
  electronic background intensity. }
\label{ResonantExcProf}
\end{figure*}

Raman scattering experiments have shown that the frequency of the \E2g mode 
in single crystals at room temperature is around 
79\,meV\cite{QuiltyPRL88,RenkerJLTP} whereas theoretical calculations
systematically underestimate this value by about
10\,meV\cite{Yildirim,Shukla}. 
It has been suggested that if the
\E2g band around the $\Gamma$-point is anharmonic then the \E2g mode
frequency is increased by the missing amount to match the
experimentally observed value\cite{Liu,Boeri}. 
In addition, the experimentally observed $T_c$ and the reduced isotope 
effect\cite{HinksPhC} can only be reconciled within anisotropic 
strong coupling theory if the \E2g mode anharmonicity is explicitly 
included\cite{Choi,ChoiPhysRevB}.  

Since the \E2g mode is responsible for a significant part of the
SC pairing a renormalization of the mode frequency upon entering 
into the SC phase is expected.
Moreover, it was predicted that the mode renormalization along the 
$\Gamma - {\rm A}$ direction of the BZ is particularly strong
\cite{Liu}.
The temperature dependence of phonon dispersions has been studied by
inelastic x-ray scattering, however, the expected self-energy
effects have not been demonstrated \cite{Shukla,Baron}. In earlier
Raman work on single crystalline \MgB2 a change of line width
between room temperature and $T_c$ has been observed
\cite{QuiltyPRL88} but not discussed. In the experimental context
it is important to notice that single crystal studies should
preferably be considered as a valid gauge of \MgB2 bulk properties
because results from powdered, polycrystalline, ceramic, film,
wire etc. samples show significant variations, especially in relation to 
phonon excitations (compare Refs.\cite{Goncharov,Martinho}). 

In this work we provide spectroscopic information on 
(i) the \E2g phonon excitation resonance, 
(ii) the reasons for the mode's large linewidth, and 
(iii) the self-energy effects upon entering into the SC state.
In particular,
(i) by means of the resonant Raman excitation profile (RREP) we show that
the light couples to the \E2g phonon \emph{via} $\sigma - \pi$ inter-band
transition;
(ii) we study the \E2g phonon shape as a function of
temperature and fit the linewidth temperature dependence with an
anharmonic decay model; (iii) we observe the \E2g phonon
renormalization (mode hardening $\sim 2.5\%$) upon entering into
the SC phase.

Raman scattering from the $ab$ surface of \MgB2 single
crystals grown as described in Ref.\cite{Karpinski} was performed in
backscattering geometry using about 1\,mW of incident power focused
to a $100 \times 200\, \mu$m spot. We have collected data from two
samples that are denoted as crystal ${\mathcal A}$ and crystal
${\mathcal B}$ when a distinction is necessary. 
The magnetic field data were acquired with a continuous flow cryostat
inserted into the horizontal bore of a SC magnet. The sample
temperatures quoted have been corrected for laser heating. 
We used the 
excitation lines of a Kr$^{+}$ laser and a triple-grating
spectrometer for analysis of the scattered light. The data were
corrected for the spectral response of the spectrometer and CCD
detector and for optical properties of the material at different
wavelengths as described in Ref. \cite{Blumberg94}.

The point group associated with \MgB2 is $D_{6h}$; subsequently, 
polarized Raman scattering from the $ab$ surface couples to two
symmetry channels: $E_{2g}$ and $A_{1g}$. The corresponding
symmetry channels can be resolved if the appropriate light
polarization is chosen for the in- and out-coming photons; in
this work only the \E2g symmetry channel will be discussed. We
denote by $(\textbf{e}_{in} \textbf{e}_{out})$ a configuration in
which the incoming/outgoing photons are polarized along the
$\textbf{e}_{in}$/$\textbf{e}_{out}$ directions. In the $VH$
configuration the light is linearly polarized with vertical ($V$)
or horizontal ($H$) directions chosen perpendicular or parallel to
the crystallographic $a$-axis. The ''right-left'' ($RL$) notation
refers to circular polarization: $\textbf{e}_{in} = (H - i V) /
\sqrt{2}$, with $\textbf{e}_{out} = \textbf{e}_{in}^{*}$ for the
$RL$ geometry. Both the $RL$ and the $V\!H$ polarization
combinations select the $E_{2g}$ representation.

\begin{figure}[t]
\includegraphics[width=0.8\columnwidth]{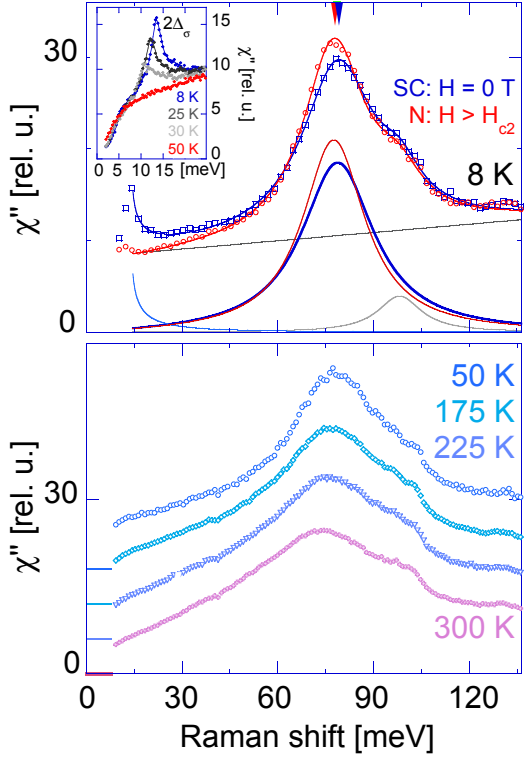}
\caption{(Color online) 
  Raman response function in the \E2g channel for $RL$ polarization 
  and 2.6\,eV excitation.
  Panel (a): Raman response from crystal ${\mathcal B}$ at 8\,K.
  Spectra at zero field and 7\,T are compared. 
  Spectra are decomposed into two Raman oscillators, 
  electronic background (linear slope), and coherence peak 
  singularity for the SC state, 
  $\sim [\omega \, \sqrt{ \left(\omega^2 - (2\Delta)^2 
  \right)}]^{-1}$.
  The arrows at the top  indicate the shift of the
  \E2g mode in the transition from the SC state (blue squares) into the 
  normal state (red circles).
  Inset: evolution of low-frequency Raman response with temperature
  across the SC transition.
  Panel (b): Data from crystal ${\mathcal A}$.
  Temperature dependence of the \E2g phonon mode above $T_{c}$. 
  Fit parameters to these spectra
  are displayed in Fig.\,\ref{fig:E2gFrequencyAndLinewidtTempDep}.}
\label{fig:Renorm}
\end{figure}
In Fig.\,\ref{ResonantExcProf} we show the Raman response as a function
of excitation energy
sampling laser lines $\lambda_{L} = 752.5, 647.1, 482.5$, and 
$406.7\,{\rm nm}$. 
We chose steps in excitation
wavelengths to resolve the broad band at around 2.6\,eV in the \MgB2
optical conductivity \cite{Guritanu}.

\begin{figure}
\includegraphics[width=0.8\columnwidth]{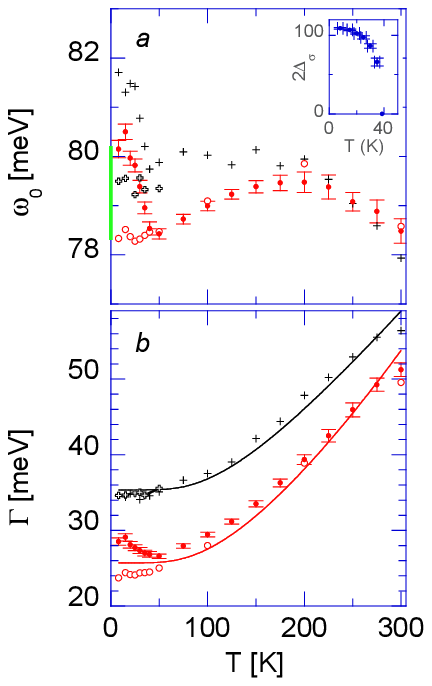}\\
  \caption{(Color online) 
Temperature dependences for the \E2g phonon frequency, $\omega_0(T)$,
(\emph{a}) and damping constants, $\Gamma(T)$, (\emph{b}) for two
\MgB2 single crystals $\mathcal A$ (black crosses) and $\mathcal
B$ (red circles) for zero field cooling (solid symbols) and
field cooling at 8~T$>H_{c2}$ (empty symbols). 
Solid lines
are fits of damping constants $\Gamma(T)$ in the normal state with
eq.~(\ref{damping}). 
The parameters from the anharmonic decay fits
for both crystals are summarized in Table\,\ref{tab:AnhDecay}. 
Inset shows temperature
dependence of the $2\Delta_{\sigma}$ coherence peak which follows
from the evaluation of spectra shown in the inset to
Fig.\,\ref{fig:Renorm}\,\emph{a}. 
The green bar at the $y$-axis
indicates the \E2g phonon frequency renormalization upon
the SC phase transition.}
\label{fig:E2gFrequencyAndLinewidtTempDep}
\end{figure}

Typical Raman features of \E2g symmetry from a \MgB2 single
crystal are shown in Fig.\,\ref{ResonantExcProf}\,\emph{c}
where phonon modes are designated with arrows.
We see that the \E2g mode at $\sim 80$\,meV (\emph{I.}) dominates the
Raman response.
In addition two-phonon scattering peaks \emph{II.} and \emph{III.}
corresponding to flat portions of the phonon dispersion curves at half 
the frequency of the respective peaks are also observed 
\cite{Shukla,footnote}.  
Finally the
spectra exhibit a strong electronic continuum with a SC pair breaking
peak at 13.6\,meV corresponding to twice the value of the SC gap
in the $\sigma$ bands \cite{QuiltyPRL88}. 
The evolution of the pair breaking peak with cooling below $T_{c}$ is 
shown in insets of Figs.\,\ref{fig:Renorm}\,\emph{a} and 
\ref{fig:E2gFrequencyAndLinewidtTempDep}\,\emph{a}.  

The RREP relates Raman scattering intensity to the interband 
transitions seen in the optical conductivity $\sigma(\omega)$.
It is particularly helpful in the
visible range because the interband contribution to 
$\sigma_{ab}(\omega)$ contains a pronounced band
around 2.6~eV \cite{Kuz'menko,Guritanu}. 
In Fig.\,\ref{ResonantExcProf}\,\emph{a-d} we show that the 
intensity of Raman response is in resonance with the 2.6\,eV band
(Fig.\,\ref{ResonantExcProf}\,\emph{e}).
We note that the ratio of the \E2g mode intensity to the strength of
the electronic background is also largest
around the 2.6\,eV excitation.
The Raman intensity of the \E2g mode is
plotted along the right y-axis in
Fig.\,\ref{ResonantExcProf}\,\emph{e} as a function of excitation
while the mode intensity to background ratio is represented by the size of
the data symbols in the plot.
The 2.6\,eV peak in optical conductivity is associated with the $\pi
\rightarrow \sigma$ electronic transitions in the vicinity of the
$\Gamma$ point and $\sigma \rightarrow \pi$ transitions in the
vicinity of the $M$ point of the BZ
\cite{Antropov,Mazin,KortusPRL86Sc_of_metallicBinMgB2}. The
RREP indicates that coupling to the \E2g mode is realized via the 
$\pi \leftrightarrow \sigma$ interband transition. 
The strongly resonant behavior of
the \E2g mode places experimental restrictions on how this mode
should be investigated.

In Fig.\,\ref{fig:Renorm}\,\emph{a} Raman spectra for the SC state
(zero field) and normal state ($H>H_{c2}$) are juxtaposed,
demonstrating the effect of magnetic field applied parallel to the 
\emph{c}-axis at 8\,K. The \E2g phonon mode hardens below the SC
phase transition along with a slight renormalization of its 
linewidth.  
At the same time a SC pair breaking peak at
13.6\,meV appears indicating the presence of the SC gap. In
Fig.\,\ref{fig:Renorm}\,\emph{b} we demonstrate how the linewidth
of the \E2g mode monotonically narrows from room temperature down
to \Tc. To quantify the dependence of the \E2g phonon frequency
and linewidth as a function of temperature and magnetic field  we
fitted the data with a sum of phononic oscillators
and a SC pair breaking singularity
on an electronic background as demonstrated in Fig.\,\ref{fig:Renorm}\emph{a}.

In Fig.\,\ref{fig:E2gFrequencyAndLinewidtTempDep} we evaluate the
\E2g phonon frequency $\omega_0(T)$ and the damping constant
$\Gamma(T)$ for both crystals ${\mathcal A}$ and ${\mathcal B}$
where we distinguish between the respective values for the normal 
and SC states. 
The solid line in Fig.\,3\,\emph{b} is a fit of the
damping constant $\Gamma(T)$ in the normal state to a model of
anharmonic two and three phonon decay at one-half and one-third
frequencies:
\begin{eqnarray}
    \Gamma(T) & = & \Gamma_{0} +
\Gamma_{3}[1 + 2n(\Omega(T)/2)] + \nonumber \\
& & \Gamma_{4}[1 +
3n(\Omega(T)/3)+3n^2(\Omega(T)/3)].
\label{damping}
\end{eqnarray}
Here $\Omega(T)={h\,c\, \omega_{h}}/{k_{B}T}$, with the harmonic
frequency $\omega_{h} = 67$\,meV
\cite{Shukla,Mazin,KortusPRL86Sc_of_metallicBinMgB2}, $n(x)$ is
the Bose-Einstein distribution function, $\Gamma_{0}$ is the
internal temperature independent linewidth of the phonon, and
$\Gamma_{3,4}$ are broadening coefficients due to the cubic and
quartic anharmonicity. The results of the fit to this anharmonic
decay model are collected in Table \ref{tab:AnhDecay}. 
For both
crystals the broadening coefficients $\Gamma_{3} + \Gamma_{4} \gg
\Gamma_{0}$ and therefore the anharmonic decay is primarily
responsible for the large damping constant of the \E2g phonon. 
We identify the reason for this rapid phononic decay in the phononic
density of states (PDOS) peaking at 33\,meV, half of the harmonic
\E2g phonon frequency $\omega_h$ (see Fig-s,1\,\emph{a} in Ref.
\cite{Yildirim} and Ref.\cite{Osborn}), which corresponds to the Van-Hove 
singularity of the lower acoustic branch 
(almost dispersionless along the $\Gamma-{\rm K}-{\rm M}$ direction, see 
Fig.\,3 in Ref.\cite{Bohnen}). 
In this context the narrowing of the \E2g mode
with Al substitution observed in Refs.~\cite{Bohnen,RenkerJLTP} can 
be readily explained with the \E2g phonon branch moving to energies 
above 100\,meV with increased Al concentration
whereas the acoustic modes that provide the decay channels stay close
to their original energies with high PDOS in the energy range of 25 -
40\,meV \cite{Bohnen}.
In short, the fast decay of the \E2g mode
is due to the unique combination of its harmonic frequency
in the $\Gamma$ point (67\,meV) corresponding to high PDOS at half of 
this frequency. 
Thus two independent puzzle pieces; (i) the cubic anharmonicity
contribution being much larger than the residual linewidth and 
(ii) the phonon density peaking just at the right energy to provide
suitable decay channels, fit together, giving evidence that the
\E2g phonon anharmonicity is responsible for the unusually broad
linewidth of this mode. 
In contrast to recent theoretical
suggestions \cite{Shukla} current experiment shows that the \E2g
phonon anharmonicity in \MgB2 must not be neglected. The residual
linewidth $\Gamma_0$ that we obtain from the fit to the
anharmonic decay model, while small, is not in contradiction to
theoretical estimates \cite{Calandra,Cappelluti} of the
electron-phonon decay contribution to the \E2g phonon linewidth.
\begin{table}[t]
\caption{
Comparison of $T_{c}$ and the $E_{2g}$ oscillator parameters for
crystals ${\mathcal A}$ and ${\mathcal B}$.}
\begin{tabular}{cccccccc}
\hline
   crystal & $T_{c}$\tablenote{in K, from SQUID measurements at~2~G;
$^{b}$in meV}
   & $\omega_0^{Nb}$    & $\omega_0^{SCb}$ & $\Gamma_{0}^{b}$ &
   $\Gamma_{3}^{b}$ & $\Gamma_{4}^{b}$ & $\kappa$ (\%) \\
\hline\hline ${\mathcal A}$ & 38.2 & 79.3 & 81.7 &
$4.0\!\pm\!1.5$ &
$31.4\!\pm\!1.2$ & small & $2.8\!\pm\!0.5$\\
${\mathcal B}$ & 38.5 & 78.1 & 80.5 & small &
$22.9\!\pm\!0.7$ &
$2.8\!\pm\!0.4$ & $2.3\!\pm\!0.3$\\
\hline
\end{tabular}
\label{tab:AnhDecay}
\end{table}

It is worth noting that individual $\Gamma_i$ parameters differ for 
the two single crystals despite the fact that both
samples were grown in the same batch.
The \E2g mode for crystal ${\mathcal A}$ is by about
10\,meV broader than for crystal ${\mathcal B}$.
With $\Gamma_0^{\mathcal A}$
somewhat higher than $\Gamma_0^{\mathcal B}$ and
$\Gamma_3^{\mathcal A}$ substantially higher than
$\Gamma_3^{\mathcal B}$ (see Table\,\ref{tab:AnhDecay}) the \E2g
mode in crystal ${\mathcal A}$ is more anharmonic than in crystal
${\mathcal B}$. Accordingly the crystal ${\mathcal A}$ mode is
pushed to about 1.2~meV higher frequency at low temperatures. We
note a correlation between the larger anharmonicity and slightly lower
\Tc \,in the case of crystal ${\mathcal A}$.

To describe the superconductivity induced self-energy effect we
refer to Fig.\,\ref{fig:E2gFrequencyAndLinewidtTempDep}\,\emph{a}.
Upon cooling in zero field $\omega_0(T)$ exhibits non-monotonic
but smooth behavior down to $T_c$. Then at $T_c$ it displays
abrupt hardening with $\omega_0^{SC}(T)$ scaling to the functional
form of the SC gap magnitude $2\Delta_{\sigma}(T)$
(Fig.\,\ref{fig:E2gFrequencyAndLinewidtTempDep}\,inset). For
in-field cooling the \E2g phonon frequency $\omega_0^N(T)$ remains
unrenormalized. The differences between the phonon frequencies in
the normal and SC states at 8~K are
$2.2 \pm 0.4$ and $1.8 \pm 0.2$\,meV for crystals  ${\mathcal A}$
and ${\mathcal B}$ respectively.
To quantify the relative hardening of the \E2g mode
we obtain the superconductivity induced renormalization
constant $\kappa = (\omega_0^{SC}/{\omega_0^{N}}) - 1 \approx 2.5\%$ 
(see Table\,\ref{tab:AnhDecay}) which 
is much smaller than the theoretically predicted
$\kappa \approx 12$\% \cite{Liu}.

We estimate the electron-phonon coupling constant
$\lambda^{\Gamma}_{E_{2g}}$ around the BZ center using
approximations adopted in Refs.~\cite{Zeyher,Rodriguez}: $\lambda
= -\kappa \, {\mathcal Re}\,(\frac{\sin u}{u})$, where $u \equiv
\pi + 2 i \cosh^{-1}({\omega^{N}}/{2\Delta_{\sigma}})$, and obtain
$\lambda^{\Gamma}_{E_{2g}} \approx 0.3$, which  
is smaller than the theoretically
predicted values (compare $\lambda$ up to unity in 
Refs.~\cite{Liu,KortusPRL86Sc_of_metallicBinMgB2,Golubov}).

In summary, we have measured the polarization resolved Raman response
as a function of excitation energy, 
temperature, and field for \MgB2 single crystals. 
From the
temperature dependence of the \E2g boron stretching phonon we
conclude that anharmonic decay is primarily responsible for the
anomalously large damping constant of this mode.
For this phonon we observe a SC induced self-energy effect and
estimate the electron-phonon coupling constant.

We acknowledge V.~Guritanu and A.~Kuz'menko for providing optical
data, and I.~Mazin and W.\,E.~Pickett for valuable discussions. 
A.\,M. has been supported in part by the German National
Academic Foundation and by the Lucent-Rutgers Fellowship Program.

\end{document}